\documentclass[preprint, twocolumn,reprint,
 amsmath,amssymb,
 aps, physrev,
]{revtex4-2}
\usepackage{subcaption}
\usepackage{graphicx}
\usepackage{dcolumn}
\usepackage{bm}

\usepackage{diagbox}
\usepackage{makecell}

\begin{document}

\preprint{APS/123-QED}

\title{\textbf{Trap-induced atom-ion complexes: a time-independent approach} 
}%

\author{Zhongqi Liang}
 \author{Ruiren Shi}
\author{Jes\'us P\'erez-R\'ios}%
 \email{Contact author: jesus.perezrios@stonybrook.edu}
\affiliation{%
Department of Physics and Astronomy, Stony Brook University, 11794 Stony Brook, NY, USA
}%

\date{\today}

\begin{abstract}

A trapped ion immersed in a neutral bath shows long-lived atom-ion complexes that significantly alter its chemical properties, and, thus the ion stability. In this work, we present a general study of trapped ion-atom scattering with the ion modeled as a charge distribution defined by the spatial extent of its ground-state wavefunction. After mapping the time-dependent problem onto a time-independent framework, we investigate the role of the trap, the atomic species, atom-ion interaction, and collision energy in shaping the chaotic dynamics of the system. We find that the probability of atom-ion complex formation directly measures its chaoticity. Therefore, our results establish a clear relationship between the emergence of chaotic scattering and the presence of ion-atom complexes.

\end{abstract}

\maketitle


\section{Introduction}

Many chemical reactions proceed through or are facilitated by the formation of a transient complex, i.e., a quasi-bound state between reactants that eventually, under suitable conditions, will transform into products. For instance, in termolecular reactions, the Lindemann-Hinshelwood mechanism describes how two reactants can first form a complex that is then stabilized by a third body~\cite{Review3BR}. Similarly, in unimolecular reactions, the Rice-Ramsperger-Kassel-Marcus (RRKM) theory relies on the existence of an activated complex that subsequently decays into product states~\cite{Marcus,rrkm1,rrkm2}. For most room-temperature chemical reactions, such complexes are short-lived in comparison with the typical duration of an experimental run, rendering them hardly detectable. In contrast, in the ultraocold regime, these complexes can be long-lived and their effects on the stability of ultracold matter are readily observable~\cite{Mayle2012,Mayle2013,ChristianenPRA,observation_bis,observationRbCs,observationHamburg,observation}. 

Long-lived complexes may also arise in ion-atom hybrid experiments, where the trapping potential confining the ion induces the formation of atom-ion complexes~\cite{Cetina_2012}. These complexes could potentially enhance ion losses via three-body recombination~\cite{Henrik2023}, or affect the chemical properties of the ion, as has been experimentally observed through enhancement in spin-exchange reactions~\cite{Ozeri2023}. Moreover, recently, it has been shown that the scattering between a free atom and a trapped ion is chaotic, suggesting signatures of quantum chaos in the ultracold regime~\cite{pinkas2024chaoticscatteringultracoldatomion}. However, despite these efforts, there is still no explanation of the origin of chaotic scattering for ion-atom systems and, more importantly, of which physical properties of the system are the most relevant to its dynamics. In addition, a connection between ion heating and ion-atom complex formation has yet to be established.

Previous studies on ion-atom scattering typically model the ion as a point charge interacting with a neutral atom, producing the expected charge-induced-dipole interaction that scales as $\propto r^{-4}$, where $r$ is the ion-atom distance~\cite{COTE20166,Perez-Rios2020,LOUS202265,RevModPhysatomion}. However, it is possible to bring the ion to the ground state of the trapping potential through side-band cooling~\cite{Michael2012} in which case the ion can be treated as a delocalized charged particle. Then the ion, instead of a point charge, can be described by a charge distribution given by its ground-state wavefunction inside the trapping potential. This allows us to map a trapped ion in a time-dependent potential into a static charge distribution with a spatial extent -- enabling a fully time-independent treatment of ion-atom scattering, as shown in Ref.~\cite{shi2025effectsdelocalizedchargedistribution}.

\begin{figure*}[t]
    \includegraphics[width=1\linewidth]{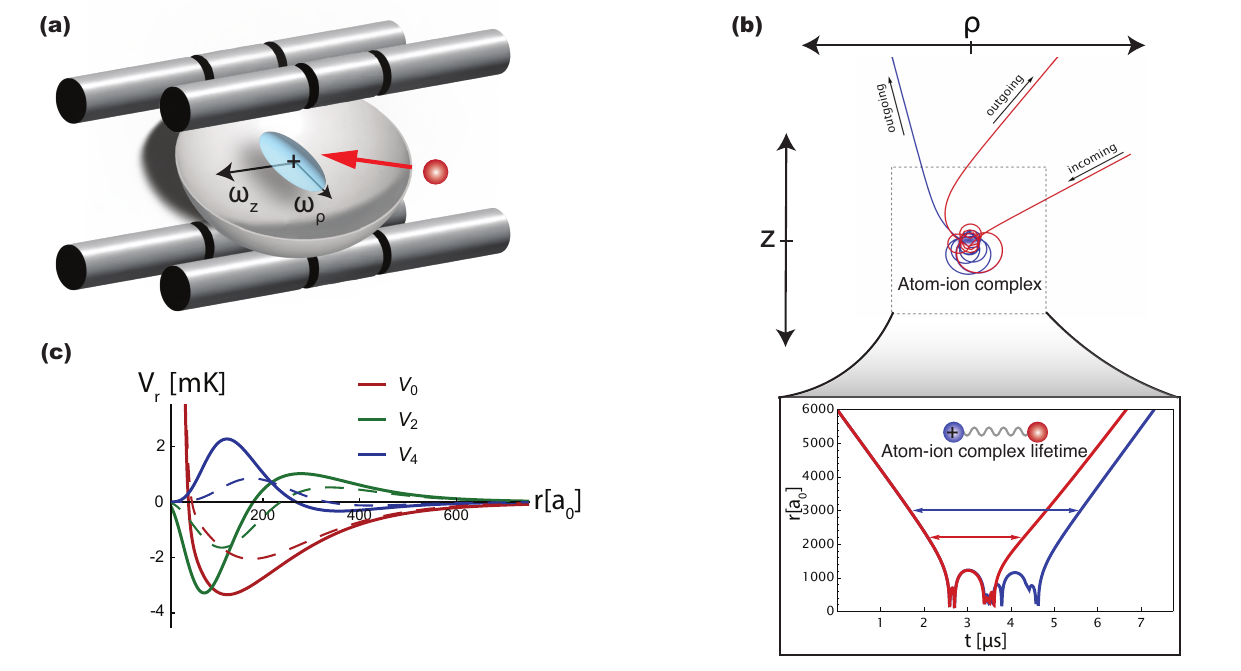}
    \caption{Trapped ion-atom interaction and scattering. Panel (a) displays a schematic representation of a trapped ion colliding with an atom. Panel (b) shows two distinct atomic trajectories for very similar values of the initial angle solved for the potential including the first two anisotropic terms. The zoom-in shows the atom-ion complex lifetime. Panel (c) displays the trapped ion-atom interaction potential. The different colors denote the three main terms of the potential expansion following Eq.~(\ref{potential_expansion}) for $\omega_z$ = 1 MHz and $\omega_\rho$ = 50 kHz (solid lines), and  for $\omega_z$ = 350 kHz and $\omega_\rho$ = 50 kHz (dashed lines).}
    \label{fig1}
\end{figure*}

In this paper, we present a detailed study of the dynamics of a trapped ion colliding with an atom, revealing the nature of the onset of chaotic scattering. In our approach, instead of treating the ion as a point charge in an external time-dependent potential, as suggested in Ref.~\cite{shi2025effectsdelocalizedchargedistribution} and illustrated in panel (a) of Fig.~\ref{fig1}. We use the delay time--the time it spends a trajectory within the interaction region--as the relevant scattering observable to investigate the effect of the short-range physics, anisotropy, trap, and atomic species in detail. Our results show that the most relevant parameters are the atomic species and the trap properties and that the chaos in the trapped ion-atom dynamics system is hyperbolic. Finally, we show that the probability of ion-atom complex formation can be viewed as a metric of chaos in the dynamical system, thus establishing a direct link between an observable quantity and an intrinsic property in dynamical systems.


\section{Hamiltonian of the Trapped Ion - Atom System}

\subsection{Ionic charge distribution}

We consider atom-ion collisions in the presence of a Paul trap. The ion is assumed to be in the ground state of the trapping potential, characterized by the radial and axial frequencies $\omega_\rho$ and $\omega_z$, respectively, as shown in panel (a) of Fig.~\ref{fig1}. In this scenario, the atom interacting with the ion will see a charge distribution determined by the spatial extent of the ion's ground-state wavefunction as~\cite{shi2025effectsdelocalizedchargedistribution}

\begin{align}
\label{eq1}
\varrho(\boldsymbol{r}) &= e|\Psi_0(\boldsymbol{r})|^2 \notag \\
&= e \frac{m \omega_\rho}{\pi} \sqrt{\frac{m \omega_z}{\pi}} \exp(-m\omega_\rho \rho^2 - m\omega_z z^2),    
\end{align}
where $m$ is the ion mass, $e$ is the electron charge and $\Psi_{0}(\boldsymbol{r})$ is the wavefunction of the ground state of the ion in the Paul trap. The charge distribution is expressed in cylindrical coordinates $(\rho,z)$  to better reflect its symmetry. Then, the interaction potential between the atom and the trapped ion is given by 

\begin{equation}
\label{eq2}
        V(r, \theta) = \frac{C_8}{r^8} - \frac{1}{2} \alpha |\boldsymbol{E} (r, \theta)|^2
    \end{equation}
where $C_8$ is the parameter that controls short-range interaction, $\alpha$ is the polarizability of the atom and $\boldsymbol{E}(r, \theta)$ is the electric field created by the charge distribution [Eq.~(\ref{eq1})]. 

\subsection{Calculation of the electric field}

Due to the symmetry of the charge distribution, it is preferable to calculate static quantities in cylindrical coordinates, in which the electrostatic potential of the charge distribution is given by
\begin{equation}
    V_e (\boldsymbol{r}) = \int \frac{\varrho(\boldsymbol{r}_0)}{|\boldsymbol{r}-\boldsymbol{r}_0|} d^3 \boldsymbol{r}_0
\end{equation}
where 
\begin{equation}
    |\boldsymbol{r}-\boldsymbol{r}_0|^2 = \rho^2 + \rho_0^2 - 2 \rho \rho_0 \cos (\phi - \phi_0) + (z - z_0)^2.
\end{equation}
Owing to the cylindrical symmetry of the trap configuration, $V_e (\boldsymbol{r})$, hence $\boldsymbol{E}(\boldsymbol{r})$, is independent of the azimuthal angle $\phi$. Therefore, without loss of generality, we may set $\phi = 0$ for simplification.

The electric field components are obtained by taking appropriate partial derivatives of the potential under the integral sign, resulting in
\begin{align}
    E_\rho (\rho,z) &= \int \frac{\varrho(\boldsymbol{r}_0)}{|\boldsymbol{r}-\boldsymbol{r}_0|^3} (\rho-\rho_0 \cos \phi_0) \, d^3 \boldsymbol{r}_0, \\
    E_z (\rho,z) &=  \int \frac{\varrho(\boldsymbol{r}_0)}{|\boldsymbol{r}-\boldsymbol{r}_0|^3} (z-z_0) \, d^3 \boldsymbol{r}_0,
\end{align}
and the field strength is
\begin{equation}
    |\boldsymbol{E}(\rho,z)|^2 = E_\rho^2 (\rho,z) + E_z^2 (\rho,z).
\end{equation}

However, for dynamics, the more natural variables are the interparticle distance $r$ and the polar angle $\theta$. Therefore, we will use $|\boldsymbol{E}(r,\theta)|^2$ rather than $|\boldsymbol{E}(\rho,z)|^2$ for later simulations. We compute $|\boldsymbol{E}(r,\theta)|^2$ numerically at 373 values of the radial coordinate $r$ ranging from 0.5~a$_0$ to 4000~a$_0$, with finer spacing near the origin. For each $r$, 24 polar angles are sampled accordingly to the Gauss-Legendre quadrature scheme, which is used later in this work. Beyond $r=4000$~a$_0$, an asymptotic $r^{-4}$ dependence--corresponding to point charge-induced dipole interactions--is imposed for $|\boldsymbol{E}(r,\theta)|^2$.

\subsection{Expansion of the potential}

Based on the symmetry of the charge distribution, it is preferable to separate the radial from the angular degrees of freedom in the interaction potential as
\begin{equation}
\label{potential_expansion}
    V (r, \theta) = \sum_{l=0}^{l_{max}} V_l (r) P_l (\cos \theta),
\end{equation}
where $P_l(x)$ denotes the $l$-th order Legendre polynomials of argument $x$, $l_{max}$ is the highest-order term included in the expansion, and the radial functions $V_l(r)$ are given by
\begin{align}
    V_l (r) &= \frac{2l+1}{2} \int_0^\pi V (r, \theta)\, P_l (\cos \theta)\, \sin \theta\, d\theta \\
    &= \frac{2l+1}{2} \int_{-1}^1 V(r, \arccos x)\, P_l(x)\, dx \\
    &\approx \frac{2l+1}{2} \sum_{i=1}^{24} w_i\, V(r, \arccos x_i)\, P_l(x_i),
\label{gauss_quadrature}
\end{align}
where the last line employs the Gauss-Legendre quadrature rule with 24 sample points $x_i \in [-1, 1]$ and associated weights $w_i$. 
    
The expansion proposed [Eq.~(\ref{potential_expansion})] is characteristic of atom-molecule scattering problems, where it facilitates the simplification of coupled-channel equations and the identification of relevant anisotropic components of the interaction potential~\footnote{For instance, the $l=0$ is associated with the isotropic component of the interaction potential, $l=1$ corresponds to the dipolar component, $l=2$ to the quadrupolar term, and so forth.}. Here, due to the symmetry under $z \leftrightarrow -z$ of the charged distribution, only even values of $l$ are allowed. Moreover, since all angular dependence in the potential originates from the $|\boldsymbol{E}(r,\theta)|^2$ term, the expansion is carried out directly on this quantity.

Panel (c) of Fig.~\ref{fig1} shows $V_l(r)$ for the three lowest $l$-values in a trapped Ba$^+$-Li system, with trapping frequencies $\omega_z$ = 1 MHz and $\omega_\rho$ = 50 kHz (solid lines) and for $\omega_z$ = 350 kHz and $\omega_\rho$ = 50 kHz (dashed lines). The isotropic term, $V_{0}(r)$, shows a short-range repulsive barrier due to the $C_8/r^8$ term in the full potential $V(r,\theta)$. In contrast, the first two anisotropic terms, $V_2(r)$ and $V_{4}(r)$, approach zero at small ion-atom separations as the electric field of a charge distribution vanishes at its center. In addition, we note the effect of the trapping frequencies on the values of each of the expansion terms of the interaction potential. Therefore, the trapped ion-atom system exhibits purely long-range anisotropy. Furthermore, we find that using the potential terms shown in Fig.~\ref{fig1}(c) suffices to reproduce the full trapped ion-atom interaction with a relative root-mean-square error (RMSE) of less than 10$\%$ across all systems and trapping configurations considered in this paper~\footnote{Please note that this statement is true strictly within the range of frequencies studied in this work, beyond which one could expect more severe deviations.}.

\subsection{Solving for the delay time $t_d$}

Due to the strength of the charge-induced dipole interaction, a classical treatment of the dynamics is appropriate within the energy range relevant to cold chemistry experiments~\cite{Perez-Rios2020}. Accordingly, we simulate atomic trajectories by solving the classical Hamiltonian dynamical system defined by
\begin{equation}
\label{hamiltonian}
H=\frac{p_r^2}{2m_a}+\frac{p_\theta^2}{2m_a r^2}+V(r,\theta),
\end{equation}
where $m_a$ is the mass of the scattering atom. The corresponding equations of motion are given by 
\begin{gather}
    \dot{r} = \frac{\partial H}{\partial p_r} \hspace{1cm} \dot{p}_r = - \frac{\partial H}{\partial r} \notag \\
    \dot{\theta} = \frac{\partial H}{\partial p_\theta} \hspace{1cm} \dot{p}_\theta = - \frac{\partial H}{\partial \theta}.
\end{gather}
The potential $V(r,\theta)$ is typically expanded to include the first anisotropic term $V_2(r)$ or the first two up to $l = 4$, retrieving the total interaction potential with an accuracy better than 10\%, as explained above.

Since the interaction potential is anisotropic, the angular momentum of the system is not conserved, leaving energy the only conserved quantity in this dynamical system with two degrees of freedom. As a result, the system is not analytically integrable and may exhibit chaotic behavior. This is illustrated in panel (b) of Fig.~\ref{fig1}, where two atomic trajectories of a Li scattering off a Ba$^+$ are shown. In both cases, the atom is launched towards the ion with a collision energy of $3.158$~$\mu$K but with incoming angles $\theta_0$ that differ by $10^{-5}$ radians. However, despite their nearly identical initial conditions, the trajectories diverge significantly after several close approaches to the scattering center and ultimately leave the scattering region with dramatically different scattering angles. Moreover, the lifetimes of the resulting atom-ion complexes also differ by nearly 1~$\mu$s.

Every simulation begins with the atom located at a distance $r_0=4000$~a$_0$ away, approaching the ion with a specific incoming angle $\theta_0$, a collision energy $E_0$, and zero initial angular momentum. In this way, we ensure that the trajectory initially corresponds to a rectilinear uniform motion of the colliding partners. The full trajectory is solved numerically in Mathematica using the NDSolveValue function~\cite{Mathematica}. From the calculated trajectories, we then extract information about the delay time, defined as the time $t_d$ the atom spends inside the region specified by the ion-atom distance $r^* = 2500$~a$_0$~\cite{Seoane_2013}. This threshold is chosen to ensure that all significant interaction dynamics are captured before the atom escapes the trap. The delay time correlates with the lifetime of atom-ion complexes and hence can be directly linked to an experimental observable. For this purpose, the built-in adaptive quasi-Monte Carlo method is employed to perform an integral of the function Boole$[r(t)<r^*]$, which returns 1 when $r(t)<r^*$ and 0 otherwise, over the full simulation time~\cite{Mathematica}.

The systems explored in this work consist of a single Ba$^+$, with mass $m_{Ba} \approx 250332$ in atomic units, confined inside a Paul trap and acting as the scatterer and a neutral H, Li, or Rb approaching with a collision energy $E_0$ at various incoming angles $\theta_0$. We investigate how the dependence of $t_d$ on $\theta_0$ changes with short-range physics -- characterized by the parameter $C_8$ in the interaction potential, the dimensionless ratio between the depth of the potential $V_{\text{min}}$ and the collision energy $E_0$, the atomic species with their masses and polarizabilities presented in Table~\ref{tab:atoms}, and the anisotropy of the trap.

\begin{table}[h]
        \centering
        \begin{tabular}{ccc}
        \hline
         & Mass & Polarizability \\
         \hline
        H & 1837 & 4.507\\
        Li & 10965 & 164.11\\
        Rb & 242272 & 319.8\\
        \hline
        \hline
        \end{tabular}
        \caption{Relevant properties of H, Li, and Rb atoms in atomic units.}
        \label{tab:atoms}
\end{table}

\section{Characterizations of Chaos}

\subsection{Hyperbolicity}

In chaotic scattering systems, a key distinction is often made between hyperbolic and nonhyperbolic chaos. For nonhyperbolic chaos, there exist in the phase space of the system islands of stability, often referred to as KAM (Kolmogorov-Arnold-Moser) islands, within which trajectories remain confined indefinitely. For trajectories that start near the boundary of these islands, they can remain close for extended periods before eventually diverging. They are often called marginally stable in the literature~\cite{Meiss1992,transientChaos}. When chaos is hyperbolic, there is no stable structure as the KAM island or marginally stable dynamics within the chaotic region, and trajectories either remain close or diverge exponentially. As a result, the surviving probability, or the fraction remaining within the scattering region, of the trajectories in a hyperbolic system follows an exponential decay $P(t) \propto \exp (-t/\tau)$, while in the nonhyperbolic case, because of the existence of marginally stable orbits, the decay is slower and follows a power law -- $P(t) \propto t^{-c}$ for some $c$. The difference in their decay pattern is widely used to distinguish between these two regimes.

\begin{figure*}[t]
    \includegraphics[width=1\linewidth]{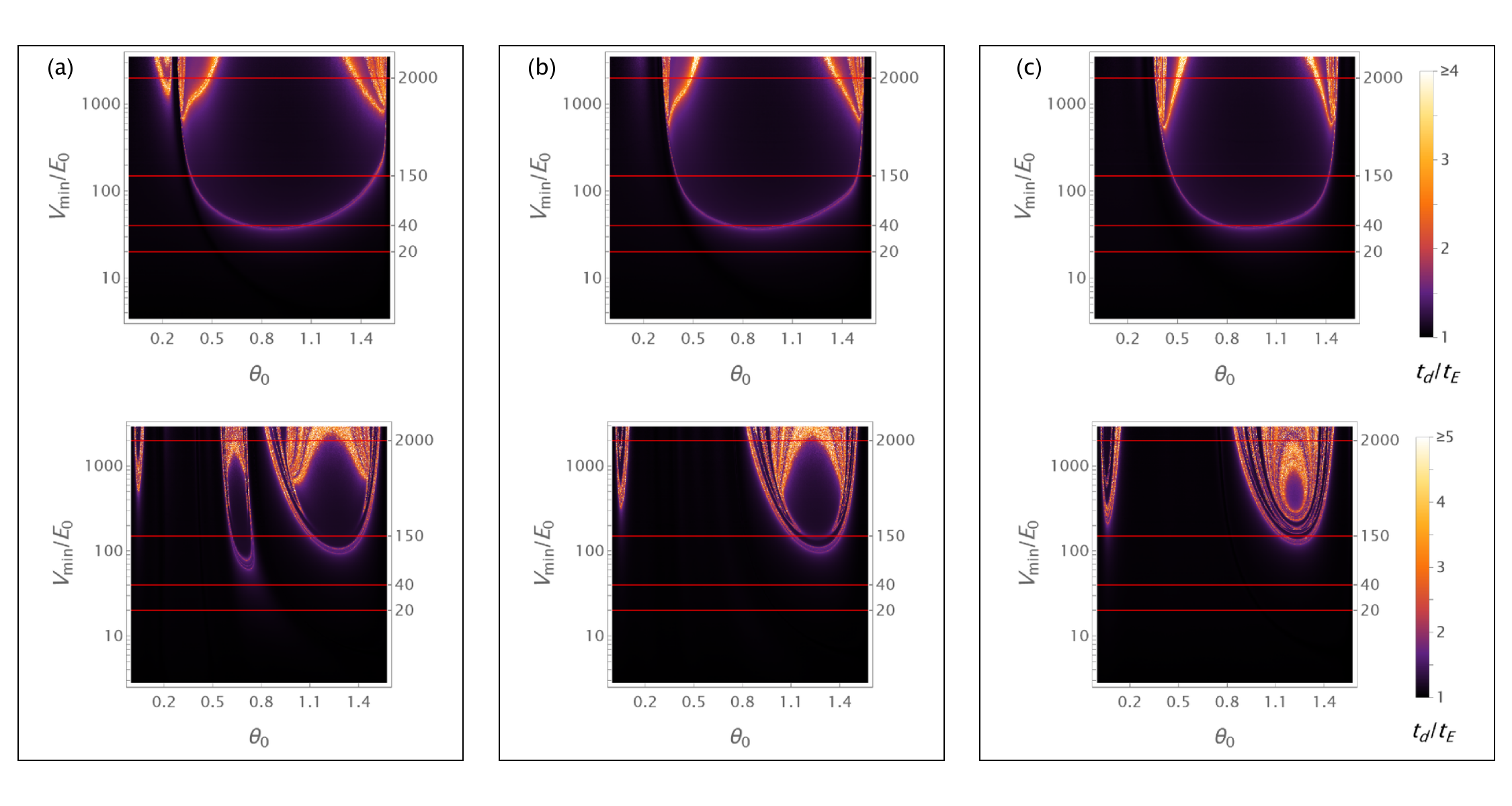}
    \caption{Delay time $t_d$ for trapped Ba$^+$-Li scattering as a function of initial angle $\theta_0$ and the ratio $V_{min}/E_0$ at $\omega_z$ = 350 kHz and $\omega_\rho$ = 50 kHz. The delay time $t_d$ is normalized against its smallest value $t_E$ at a given $E_0$ and its values are color-coded as demonstrated by the legends included in panel (c). Panel (a) has $C_8 = 10^4$~a.u., panel (b) corresponds to $C_8 = 10^5$~a.u., and $C_8 = 10^6$~a.u. for panel (c). Inside each panel, the subplot on the top solves $t_d$ with the potential only expanded to $l = 2$, whereas the one on the bottom considers the $l=4$ term as well. The horizontal red lines cut through the parameter space at $V_{min}/E_0 = 20, 40, 150, 2000$, corresponding to values for regular scattering, the onset of chaos, and deeply chaotic. A detailed look at the cuts is presented in Fig.~\ref{fig:cuts_350k50k}.}
    \label{fig:scan_350k50k}
\end{figure*}

\subsection{Chaotic Fraction and Fractal Dimension}

When the dynamics is regular, the delay time $t_d$ varies smoothly with the incoming angle $\theta_0$. However, singularities occur when nearby values of $\theta_0$ can produce $t_d$ that are drastically different. Here, we define a trajectory as unstable if a perturbation of size $\epsilon$ in $\theta_0$ produces a change in the delay time $\Delta t_d > t_E/10$, where $t_E$ is the shortest delay time observed at the given collision energy. The uncertainty fraction denoted $f(\epsilon)$, is defined as the fraction of initial conditions $\theta_0$ that produce unstable trajectories associated with the formation of atom-ion complexes under this criterion. It gives a quantitative measure of how much the parameter space exhibits chaotic behavior.

It should be noted that $f(\epsilon)$ has a power-law dependence on the resolution $\epsilon$
\begin{equation}
    f(\epsilon) \propto \epsilon^\alpha,
\end{equation}
where the exponent $\alpha$ is known as the uncertainty exponent. From this, we can define the fractal dimension $d$ as
\begin{equation}
    d = 1 - \alpha,
\end{equation}
 which characterizes the geometric complexity of the set of sensitive initial conditions. If $t_d (\theta_0)$ is a smooth function, then a smaller $\epsilon$ always corresponds to a higher predictability and $f(\epsilon)$ scales linearly with $\epsilon$, yielding $d = 0$. If there is complete unpredictability, on the other hand, then $f(\epsilon)$ is independent of $\epsilon$, which gives $d = 1$ ~\cite{Grebogi1983,Bleher1989}.

In this work, we focus on the prevalence of chaotic behavior and use the uncertainty fraction $f(\epsilon)$ at fixed resolution $\epsilon^* = 10^{-5}\times \pi/2$ as the primary measure of chaos. It is also experimentally measurable as it is equivalent to the probability of ion-atom complex formation. Therefore, it is possible to establish a link between a direct experimental observable and the onset of chaotic scattering.

\section{Results}

\subsection{Anisotropy of the Potential}


We begin by investigating how the inclusion of higher-order anisotropic terms in the interaction potential $V(r)$ influences the dynamics of the system, using the potential expansion given by Eq.~(\ref{potential_expansion}). Our results are presented in terms of the potential depth-to-collision energy ratio $V_{\text{min}}/E_0$, similar to what has been employed before in the study of chaotic scattering~\cite{ChaoticScattering}. 

With the introduction of anisotropy in the interaction potential, the angular momentum is no longer conserved, which is a necessary -- though not a sufficient -- condition for the emergence of chaos. At low collision energies $E_0$ relative to the potential depth $V_{\text{min}}$, we expect the atom to remain in the interaction region for a longer time, that is, having a longer delay time, allowing the anisotropy of the potential to significantly deflect the atom from its otherwise rectilinear path, potentially leading to atom-ion complex formation.

To examine in more detail how the delay time $t_d$ depends on the ratio $V_{\text{min}}/E_0$ and the initial condition $\theta_0$, we compute $t_d$ across three orders of magnitude in the collision energy $E_0$ while sampling $\theta_0$ on the interval $[0,\pi/2]$. The lowest energy, $E_{0,\text{min}} \approx 10 \times V(r_0)$, is chosen to ensure that the atom is not initially in a bound state. A total of 500 energies are sampled, evenly spaced on a logarithmic scale. For each energy, 2000 trajectories are simulated with uniformly spaced $\theta_0$, using $\Delta \theta_0 = \pi/4000$. The resulting maps of delay time are shown in Fig.~\ref{fig:scan_350k50k}, where fractal structures, symptomatic of chaotic scattering, emerge at high $V_{\text{min}}/E_0$ values. The upper subplots includes $V_{2}(r)$, whereas the lower panes include $V_{2}(r)$ and $V_{4}(r)$ terms in the potential expansion. Independently of the number of anisotropic terms in the potential expansion, a significant fraction of the trajectories are long-lived due to ion-atom complexes formation, and the delay time $t_d$ varies sharply with small changes in $\theta_0$, indicated by the rapid alternations between lightness and darkness and indicative of chaotic scattering. However, with the second anisotropic term included, the angular dependence of the interaction becomes more complex, and more intricate and densely packed structures emerge in the parameter space, suggesting a more chaotic landscape. 

\begin{figure}[h]
    \centering
    \includegraphics[width=1\linewidth]{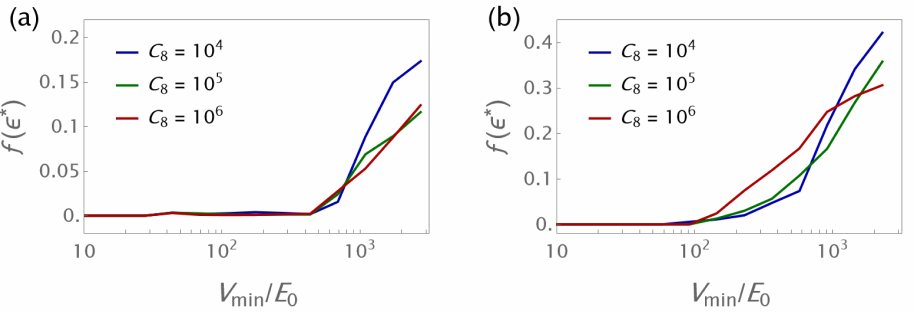}
    \caption{Uncertainty fraction $f(\epsilon^*)$ as a function of $V_{\text{min}}/E_0$ for a Li scattering off a Ba$^+$ inside an ion trap that has $\omega_z$ = 350 kHz and $\omega_\rho$ = 50 kHz for different values of $C_8$ (in a.u.). Panel (a) is calculated when only $V_0(r)$ and $V_2(r)$ are included, whereas panel (b) also takes into account $V_4(r)$.}
    \label{fig:unc_frac_diff_c8}
\end{figure}

A different route to address the effect of the anisotropy in the onset of chaotic scattering is to calculate the uncertainty fraction $f(\epsilon^*)$ for one and two anisotropic terms, and the results are shown in panels (a) and (b) of Fig.~\ref{fig:unc_frac_diff_c8}, respectively. First, we notice a critical point in $V_{\text{min}}/E_0$, separating regular from chaotic dynamics, i.e., beyond the critical value, the uncertainty fraction stays nonzero and often grows steadily as a function of $V_{\text{min}}/E_0$. This threshold occurs at different  $V_{\text{min}}/E_0$ values depending on the number of anisotropic terms included in the dynamics. In particular, the results with more anisotropic terms show the transition to chaotic scattering at higher collision energies than the case of a single anisotropic term. However, this effect could be purely coincidental since $V_4(r)$ and $V_2(r)$ often have opposite signs, as shown in panel (c) of Fig.~\ref{fig1}, resulting in a shallower overall potential.

\begin{figure}[h]
    \centering
    \includegraphics[width=1\linewidth]{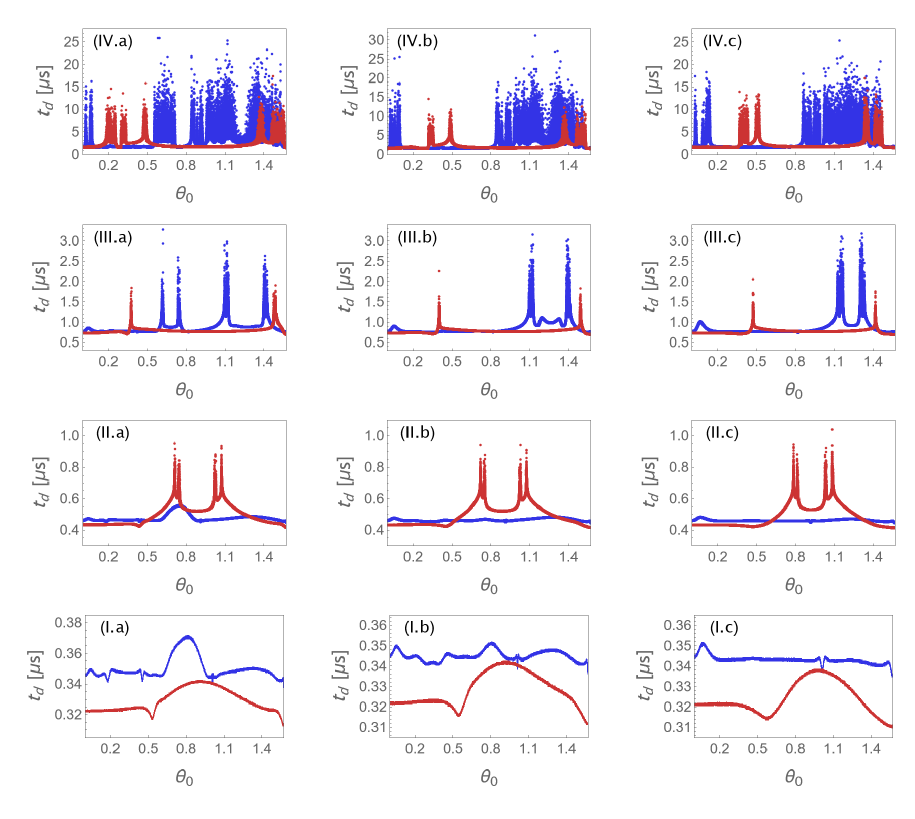}
    \caption{Delay time $t_d$ for trapped Ba$^+$-Li scattering as a function of incoming angle $\theta_0$ for $C_8 =$ (a) $10^4$~a.u., (b) $10^5$~a.u., (c) $10^6$~a.u. and $V_{\text{min}}/E_0 =$ (I) 20, (II) 40, (III) 150, (IV) 2000 at $\omega_z$ = 350 kHz and $\omega_\rho$ = 50 kH The color red is for $l= 2$ and the blue dots are for the interaction potential including terms up to $l  = 4$. The uncertainty fraction $f(\epsilon)$ at $\epsilon^* = 10^{-5} \times \pi/2$ is then calculated. Their values are shown in Table~\ref{tab:uncertainty_frac_350k50k}.}
    \label{fig:cuts_350k50k}
\end{figure}

These general observations are further supported by one-dimensional slices of $t_d(\theta_0)$ for fixed values of $V_{\text{min}}/E_0$, shown in Fig.~\ref{fig:cuts_350k50k}. These slices are sampled at the resolution $\epsilon^*$ with their corresponding uncertainty fractions $f(\epsilon^*)$ recorded in Table~\ref{tab:uncertainty_frac_350k50k}. In addition, when $V_4(r)$ is included in the potential, particularly at low energies, not only do more trajectories result in the formation of atom-ion complexes, but the complexes themselves decay more slowly and exhibit significantly longer lifetimes. These results suggest that the magnitude and the complexity of the angular dependence should be discussed separately: the former is correlated with the critical value of $V_{\text{min}}/E_0$ required for complex formation, while the latter contributes to the intricacy of the structures observed in the parameter space after surpassing the threshold.

\subsection{Short-Range Effects}


As the short-range structure of the trapped ion-atom interaction is difficult to probe in the cold and ultracold regime~\cite{astrakharchik2020ionic,Saajid2024,Henrik2022,Henrik2023}, we treat $C_8$ as a free parameter. The values considered are $C_8 = 10^4, 10^5$ and $10^6$~a.u. in this section for Ba$^+$-Li scattering in a trap with $\omega_z$ = 350 kHz and $\omega_\rho$ = 50 kHz. The corresponding potential depths, $V_{\text{min}}$, calculated when only including the first anisotropic term, are 3.4514 mK, 3.4512 mK, and 3.4486 mK, occurring at $r_{\text{min}}$ = 134.7 $a_0$, 134.8 $a_0$, and 135.0 $a_0$, respectively. When $V_4 (r)$ is taken into account, the corresponding $V_{\text{min}}$ values decrease to 2.8744 mK, 2.8747 mK, and 2.8725 mK, differing by less than 0.1 \%. The proximity of these values indicates that the long-range behavior of the system remains essentially unchanged across our chosen range of $C_8$. Therefore, any significant difference in the resulting dynamics can be attributed to purely short-range effects. 


\begin{table}[h]
    \centering
    \begin{tabular}{c c c c}
    \hline
    \diagbox{$V_{\text{min}}/E_0$}{$C_8$} & (a) $10^4$~a.u. & (b) $10^5$~a.u. & (c) $10^6$~a.u. \\
    \hline
    (I) 20 &  \makecell{0 \\ 0} & \makecell{0 \\ 0} & \makecell{0 \\ 0}\\
    (II) 40 &  \makecell{ $3.92 \times 10^{-3}$ \\ 0} & \makecell{ $3.61 \times 10^{-3}$ \\ 0} & \makecell{$4.55 \times 10^{-3}$ \\ 0} \\
    (III) 150 & \makecell{$2.15 \times 10^{-3}$ \\ 0.0119} & \makecell{$5.70 \times 10^{-4}$ \\ 0.0210} & \makecell{$5.90 \times 10^{-4}$ \\ 0.0213} \\
    (IV) 2000 & \makecell{0.153 \\ 0.408} & \makecell{0.0978 \\ 0.343} & \makecell{0.0976 \\ 0.303}\\
    \hline
    \hline
    \end{tabular}
    \caption{Uncertainty fractions for Ba$^+$-Li scattering calculated for example values of $V_{\text{min}}/E_0$ and $C_8$ at $\omega_z$ = 350 kHz and $\omega_\rho$ = 50 kHz. Inside each cell, $f(\epsilon)$ for the interaction potential up to $l = 2$ is presented on top, below which shows the value including $l = 4$.}
    \label{tab:uncertainty_frac_350k50k}
\end{table}

To study the impact of varying $C_8$, we create similar maps of delay time $t_d$ in the two-dimensional parameter space defined by $V_{\text{min}}/E_0$ and $\theta_0$, as shown in Fig.~\ref{fig:scan_350k50k}. From left to right within each row -- where the same number of terms are included in the potential expansion--as $C_8$ increases, the most apparent systematic change in the parameter space is that the region enclosed by chaotic trajectories in the top-right corner becomes progressively smaller. Slightly higher values of $V_{\text{min}}/E_0$ are required to induce chaotic motion, and the width of the interval in $\theta_0$ that contains the majority of chaotic trajectories at some given energy is also shortened. This effect is also observed in Fig.~\ref{fig:cuts_350k50k}, where each slice at different values of $V_{\text{min}}/E_0$, shows the effect of the short-range physics on the dynamics. However, within such intervals, there is no clearly discernible trend in the complexity of the dependence of the delay time on $\theta_0$ as $C_8$ varies, as demonstrated in Fig.~\ref{fig:unc_frac_diff_c8}. Despite that the smallest $C_8$, which has the deepest potential and allows the closest approach, produces the most complexes at low energies ($V_{\text{min}}/E_0>10^3)$, there does not exist a clear hierarchy for higher energies.

\begin{figure}[h]
    \centering
    \includegraphics[width=1\linewidth]{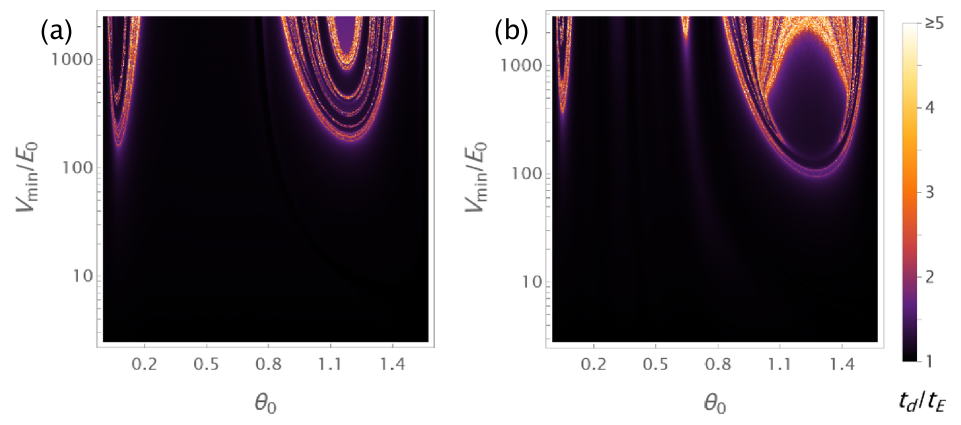}
    \caption{Delay time $t_d$ as a function of initial angle $\theta_0$ and the ratio $V_{\text{min}}/E_0$ for a H [panel (a)] or a Rb [panel (b)] scattering off a Ba$^+$ inside a Paul trap of frequencies $\omega_z$ = 350 kHz and $\omega_\rho$ = 50 kHz. The first two anisotropic terms are included in the potential to produce this figure and $C_8=10^5$~a.u..}
    \label{fig:scan_diff_atoms_350k50k}
\end{figure}

In general, the short-range interaction does play a role in the trapped ion-atom dynamics, where stronger repulsion can restrict the range in the parameter space where one can find the most chaotic trajectories. However, regarding the exact probability of complex formation, the potential with the strongest short-range repulsion does not always give the highest value for a given collision energy.

\subsection{Atomic Species}

In this section, we fix $C_8 = 10^5$~a.u. and investigate the role of atomic species in ion-atom complex formation, including $V_{2}(r)$ and $V_{4}(r)$ in the potential expansion. The trap frequencies are kept at $\omega_z$ = 350~kHz and $\omega_\rho$ = 50~kHz, so $|\boldsymbol{E}(r,\theta)|^2$ remains unchanged. In addition, the atomic mass $m$ also enters the Hamiltonian (Eq.~\ref{hamiltonian}). The atoms H, Li, and Rb are chosen because their masses $m_\text{H}$, $m_{\text{Li}}$, and $m_{\text{Rb}}$ increase in orders of magnitude. At the same time, the trend for polarizability is different -- the polarizability of Li and Rb only differ by a factor of 2 but both are approximately two orders of magnitude larger than the polarizability of hydrogen (see Table~\ref{tab:atoms}). This combination allows us to disentangle the relative importance of atomic mass and polarizability in shaping the scattering dynamics. 

\begin{figure}
    \centering
    \includegraphics[width=0.8\linewidth]{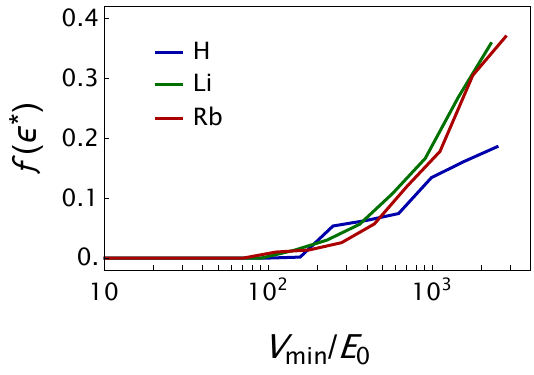}
    \caption{Uncertainty fraction $f(\epsilon^*)$ as a function of $V_{\text{min}}/E_0$ for a different atoms scattering off a Ba$^+$ inside an ion trap that has $\omega_z$ = 350 kHz and $\omega_\rho$ = 50 kHz. The short-range interaction is characterized by and $C_8=10^5$~a.u.. The delay time is calculated with the first two anisotropic terms of the potential.}
    \label{fig:unc_frac_diff_atom}
\end{figure}

The results are shown in Fig.~\ref{fig:scan_diff_atoms_350k50k}, which displays the delay time maps for H and Rb scattering off a trapped Ba$^+$. It can be observed that the corresponding maps for Ba$^+$-Li and Ba$^+$-Rb scattering exhibit very similar overall structures except for an additional chaotic region at the top of the parameter space near $\theta_0 = 0.65$ in the latter. In contrast, the ring-like pattern for Ba$^+$-H scattering observed in the upper right corner is absent in the other two cases. Furthermore, larger values of $V_{\text{min}}/E_0$ are required to trigger chaotic dynamics in the Ba$^+$-H system. These observations are consistent with Fig.~\ref{fig:unc_frac_diff_atom} where the uncertainty fraction is displayed as a function of $V_{\text{min}}/E_0$, as the curves for Li and Rb remain closely aligned and atom-ion complexes begin to form at lower values of $V_{\text{min}}/E_0$ compared to the H case. Together, these results suggest that atomic polarization, rather than atomic mass, is the dominant factor influencing atom-ion dynamics.

\subsection{Trapping Frequencies}

In addition to the intrinsic properties of the atom, the external trapping potential, by modifying the ionic wavefunction and, consequently, the charge distribution given by Eq.~(\ref{eq1}), also plays a critical role in shaping the interaction landscape. Here we vary the axial trapping frequency $\omega_z$ while keeping the radial frequency fixed at $\omega_\rho = 50$~kHz and examine the changes in the resulting dynamics. 

\begin{figure}[h]
    \centering
    \includegraphics[width=1\linewidth]{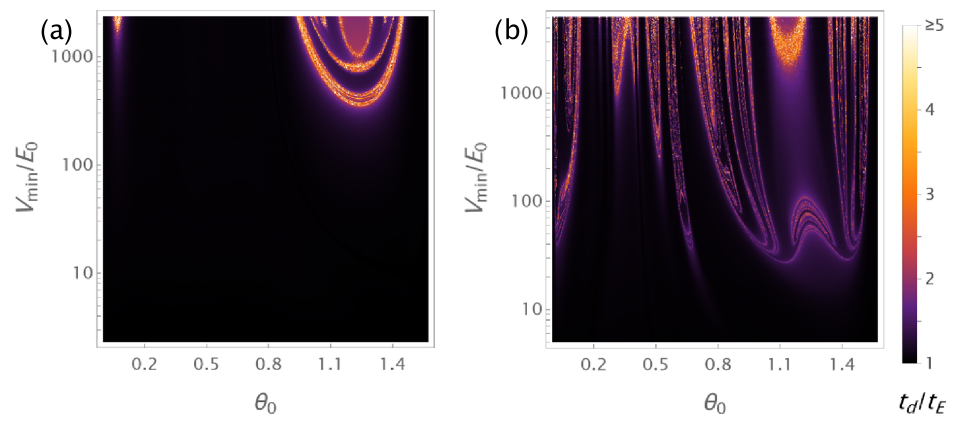}
    \caption{Delay time $t_d$ as a function of initial angle $\theta_0$ and the ratio $V_{\text{min}}/E_0$ for Ba$^+$-Li scattering at (a) $\omega_z$ = 250 kHz and (b) $\omega_z$ = 1 MHz while keeping $\omega_\rho$ = 50 kHz and $C_8=10^5$~a.u.. Terms up to $l = 4$ are included in the potential.}
    \label{fig:scan_diff_freq}
\end{figure}

\begin{figure}[h]
    \centering
    \includegraphics[width=1\linewidth]{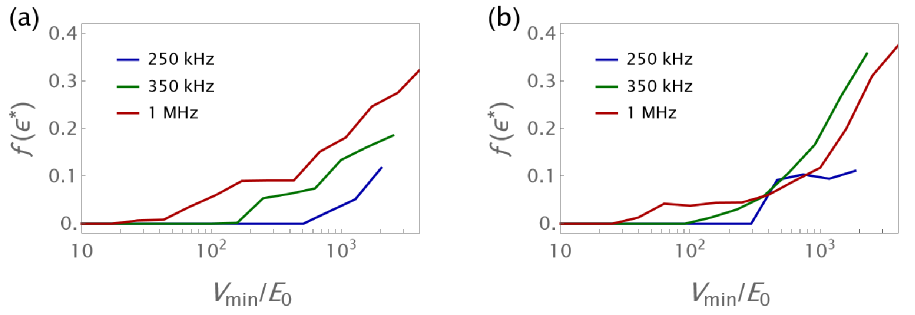}
    \caption{Uncertainty fraction $f(\epsilon^*)$ as a function of $V_{\text{min}}/E_0$ for a H [panel (a)] or Li [panel (b)] scattering off a Ba$^+$ inside an ion trap that has varied $\omega_z$ but the same $\omega_\rho$ = 50 kHz and $C_8 = 10^5$~a.u.. The delay time is calculated with the first two anisotropic terms included in the potential.}
    \label{fig:unc_frac_diff_freq}
\end{figure}

Increasing $\omega_z$ makes the charge distribution more localized in the axial direction. Therefore, angular-dependent terms such as $V_2(r)$ and $V_4(r)$ in the potential become more prominent compared to the isotropic $V_0(r)$ and shift the potential minimum $V_{\text{min}}$ closer to the trapped ion while increasing its magnitude. As a consequence, one would expect a lower threshold of $V_{\text{min}}/E_0$ for complex formation, which is indeed observed in Figs.~\ref{fig:scan_diff_freq} and~\ref{fig:unc_frac_diff_freq} for both Ba$^+$-H and Ba$^+$-Li scattering. However, whereas higher $\omega_z$ appears to lead to more complex formation at particular values of $V_{\text{min}}/E_0$ in the Ba$^+$-H system across the frequencies we considered, the same cannot be said for the Ba$^+$-Li case. Therefore, the role of the anisotropy of the trap can not be disentangled from the atomic species, which makes sense in light that the trap anisotropy establishes the relevant long-range anisotropy which the atom polarizability further enhances.

\subsection{The onset of chaotic scattering}
\begin{figure}[h]
    \centering
    \includegraphics[width=\linewidth]{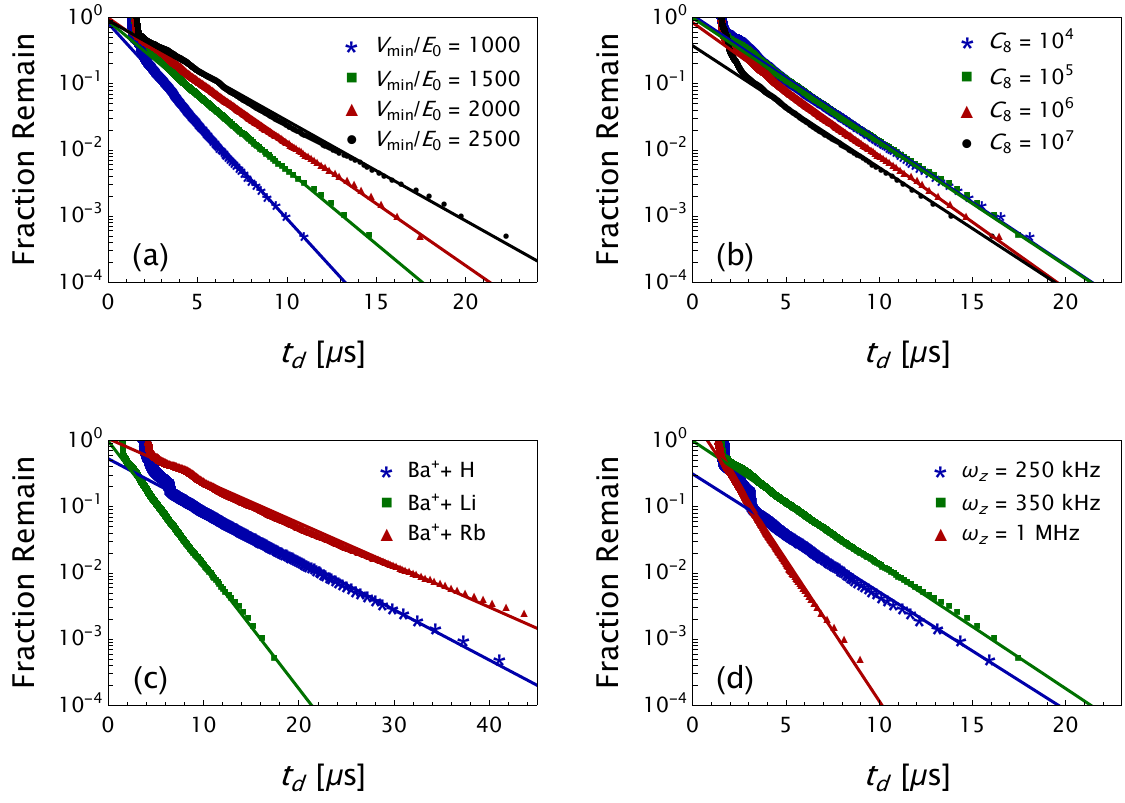}
    \caption{Fractions of atomic trajectories remaining inside the interaction region ($r < 2500 a_0$) plotted against the delay time $t_d$. The default parameters are $V_{\text{min}}/E_0$ = 2000, $C_8$ = $10^5$~a.u., and $\omega_z$ = 350 kHz for Ba$^+$-Li scattering. Inside each panel, only one parameter is varied with corresponding values specified in the legend. The first two anisotropic terms are included for the calculation.}
    \label{fig:frac_remain}
\end{figure}

We have proved that a trapped ion-atom dynamics is chaotic. However, we did not analyze the nature of chaos in the system. To do so, we compute the faction of trajectories that show a delay time larger than a given value, and the results are shown in Fig.~\ref{fig:frac_remain}, thus exploring the lifetime distribution of ion-atom complexes. Every panel of the Figure refers to a general case [panel (a)], short-range effects [panel (b)], different atoms [panel (c)] and different frequencies [panel (d)]. Despite their intricate dependence on the anisotropy of the system, the chaotic dynamics examined in this work are consistently hyperbolic. This is evidenced by the exponential decay of the complex lifetimes observed across all parameter combinations considered. This means that all the trajectories calculated in this work are unstable, and that there are no complexes formed that are marginally stable or stable with exceptionally long lifetimes.

Through Fig.~\ref{fig:frac_remain}, it is easy to see that the atomic species and the trap frequency are the two most important parameters controlling ion-atom complexes' lifetime distribution. These results result from the long-range anisotropy of the ion-atom interaction governed by the trap properties and enhanced by the polarizability of the atomic species. From panel (d), it is clear that the more anisotropic the trap is, the larger the decay rates are for the ion-atom complexes as the atoms can reach higher speeds inside the trapping region. Regarding the atomic species, it is clear that Li yields shorter-lived complexes than Rb and H, which is related to the polarizability-to-mass ratio. Finally, as shown in panel (b), the short-range effects only slightly affect ion-atom complexes' lifetime distribution.

\subsection{Limitations of the approach}

\begin{figure}[h]
    \centering
    \includegraphics[width=\linewidth]{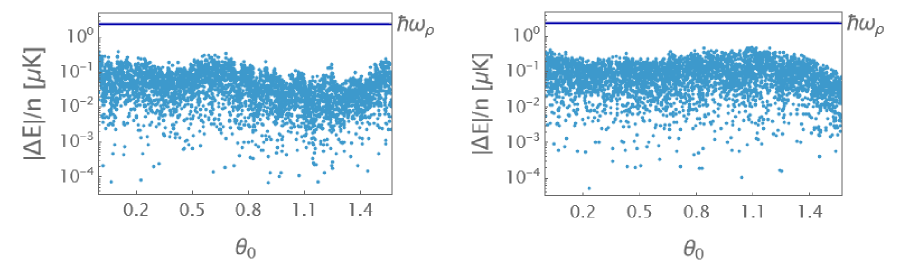}
    \caption{Average energy transfer per collision for Ba$^+$-Li scattering for 4000 random initial angles compared to the minimum energy to excite Ba$^+$ (indicated by the blue line) at (a) $V_{\text{min}}/E_0 \approx 3456$ and (b) $V_{\text{min}}/E_0 \approx 3.456$, which are the maximum and the minimum of the ratio $V_{\text{min}}/E_0$ explored in the second row of Fig.~\ref{fig:scan_350k50k} when the second anisotropic term is also included.}
    \label{fig:energy_transfer_350k50k}
\end{figure}

Our approach will only be valid when the ion wave function does not change during a collision since we assume the charge distribution is static. In other words, the average energy transfer between the ion and the atom per collision is insufficient to excite the trapped Ba$^+$ from its ground state. It, therefore, does not alter its wavefunction or the resulting charge distribution. This is illustrated in Fig.~\ref{fig:energy_transfer_350k50k}, which shows that even at its highest, the calculated energy transfer per collision remains approximately an order of magnitude less than the minimum excitation energy of the harmonic oscillator. These results confirm that our approach is suitable for simulating ion-atom collisions in the cold and ultracold regime.

On the other hand, we have not included the micromotion effects characteristic of a Paul trap since those are pretty small for the ground state of the trap. However, when considering the ion in an excited state, the micromotion effects will alter the ion wavefunction and the interaction potential that the atom feels, modifying its scattering properties.

\section{Conclusion}

In this work, we have investigated the conditions under which chaotic scattering arises and develops in trapped-ion atom collisions using classical trajectory simulations. Following  Ref.~\cite{shi2025effectsdelocalizedchargedistribution}, it is possible to substitute the trapped ion by a charge distribution dictated by the extension of the ion's wave function. Therefore, mapping the time-dependent trapped ion-atom problem into a charge distribution-atom problem, whose dynamics can be readily studied classically. We find that the onset of chaotic motion -- characterized by the lowest value of the ratio $V_{\text{min}}/E_0$ at which unstable trajectories appear -- is positively correlated with the magnitude of the trapped ion-atom interaction anisotropy, whether introduced from including different terms in the interaction potential or through experimental control of trapping frequencies. The inclusion of higher-order terms in the Legendre expansion of the interaction potential significantly complicates the structure of the chaotic regions in the parameter space. Our results also demonstrate that short-range physics has an impact on the scattering dynamics, where stronger repulsion reduces the area in the parameter space enclosed by chaotic trajectories. The role of atomic species is also explored, where we identify the atomic polarizability, rather than the atomic mass, as the dominant factor in shaping the parameter space.

Our findings indicate that lighter atoms show almost the same probability of complex formation with a trapped ion as heavier atoms, in contrast to other studies, which consider a time-dependent trap holding the ion. However, in previous studies, hydrogen as the scattering atom has never been considered ~\cite{Furst_2018,pinkas2024chaoticscatteringultracoldatomion,Ozeri2023,Pinkas_2020,Cetina_2012,Trimby_2022}. In light of these results, and our approach, it could be possible to revisit of the role of the atomic species on the ion-atom dynamics and its relevance for collision-induced heating of the trapped ion.

\section{Acknowledgments}
We thank the support Office of the Vice-President research of Stony Brook University through the Seed Program. The authors acknowledge Prof. M. Drewsen for inspiring this work, for fruitful discussions, and for reading the manuscript. 


\bibliography{apssamp}

\end{document}